\documentstyle[epsf]{elsart}

\begin{document}
\begin{frontmatter}
\title{Measurement of the thickness of an insensitive surface layer
of a PIN photodiode}
\author[UT]{Y. Akimoto\corauthref{me}},
\corauth[me]{Corresponding author.
Tel.: +81 3 5841 7622; fax: +81 3 5841 4186.}
\ead{akichan@icepp.s.u-tokyo.ac.jp}
\author[ICEPP]{Y. Inoue},
\and
\author[UT]{M. Minowa}
\address[UT]{Department of Physics, School of Science, University of Tokyo,
7-3-1, Hongo, Bunkyo-ku, Tokyo 113-0033, Japan}
\address[ICEPP]{International Center for Elementary Particle Physics, 
University of Tokyo, 7-3-1, Hongo, Bunkyo-ku,
Tokyo 113-0033, Japan}

\begin{abstract}
We measured the thickness of an insensitive surface layer of a PIN photodiode,
Hamamatsu S3590-06, used in the Tokyo Axion Helioscope. 
We made alpha-particles impinge on the PIN photodiode in 
various incidence angles and measured the pulse height
to estimate the thickness of the insensitive surface layer.
This measurement showed its thickness was $0.31 \pm 0.02 \; {\rm \mu m}$
on the assumption that the insensitive layer consisted of Si.
We calculated the peak detection efficiency for low energy
x-rays in consideration of the insensitive layer and escape of x-rays
and Auger electrons.
This result showed the efficiency for 4-10keV x-rays was more than 95\%.
 
\end{abstract}

\end{frontmatter}

\section{Introduction}
\begin{figure}[p]
\begin{center}
\epsfysize=12cm
\epsfbox{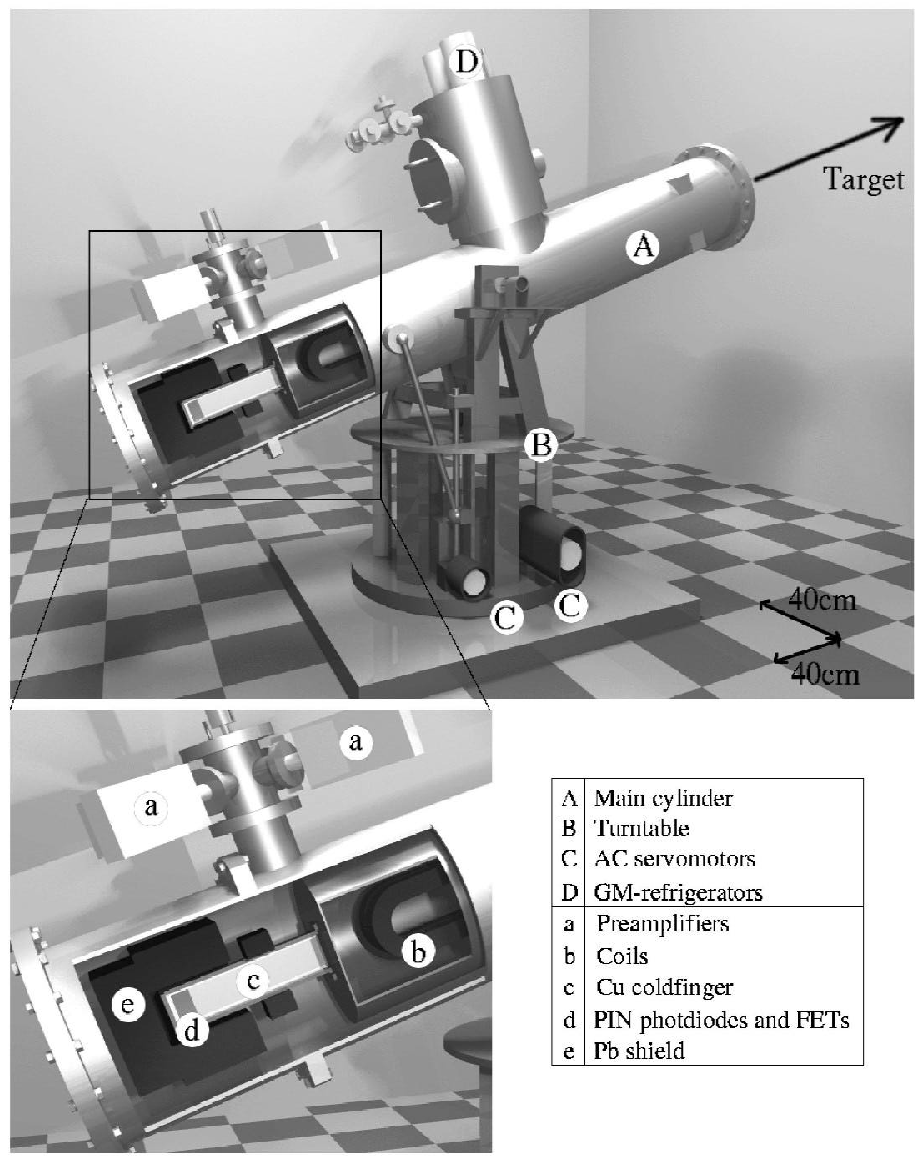}
\caption{Schematic view of the Tokyo Axion Helioscope}
\label{sumico}
\end{center}
\end{figure}

A PIN photodiode is a very useful device.  
It is used not just for a photosensor but also for a detector of
charged particles, x-rays
and gamma-rays by a lot of experiments in high energy physics.
A PIN photodiode is an improved version of the low-capacitance
planar silicon photodiode and
makes use of an extra high resistivity intrinsic silicon layer between
a P- and an N$^+$-layer to improve response time.
Reversed bias enhances response time of the PIN photodiode.
It has high resistance to breakdown
and low leakage current.
We use PIN photodiodes for an x-ray detector in the search experiment
for solar axions (the Tokyo Axion Helioscope (Fig. \ref{sumico})
 \cite{tokyo1} \cite{tokyo2}),
because of a low sensitivity to magnetic fields,
tolerance to low-temperature, and high efficiency for low energy x-rays
\cite{pin}.

However, it has a slight disadvantage.
There is non-active area so-called `a dead layer' which consists of
the SiO$_2$ layer formed on the detector surface as a protective film and
the P-layer on a PIN photodiode.
But these layers do not perform as depleted layers. They have no sensitivity.
It could worsen the sensitivity to low energy x-rays in our experiment.
In \cite{tokyo1}, we reported that the thickness was $6.1 \;{\rm \mu m}$
at 95\% confidence level.
But we thought the true value was much less than this one.

In this work, we determined the thickness of the insensitive surface layer of
the PIN photodiode
to recalculate the upper limit of axion-photon coupling constant
using the data obtained in our earlier experiments \cite{tokyo1} \cite{tokyo2}.

\section{Experimental Apparatus}
To measure the thickness of the insensitive surface layer,
we made alpha-particles from an $^{241}$Am source 
impinge on a PIN photodiode at various incidence angles.
The range for the 5.486 MeV alpha-particle in silicon is ${24 \; {\rm \mu m}}$
and much less than the thickness of the expected depleted layer.
Respective path lengths of alpha-particles in the insensitive and sensitive
layers depend on their incidence angle.
Therefore, it enables us to estimate the thickness of the insensitive layer
measuring pulse heights of alpha-particles in the PIN photodiode
as a function of the incidence angle.

\begin{figure}[p]
\begin{center}
\epsfxsize=8cm
\epsfbox{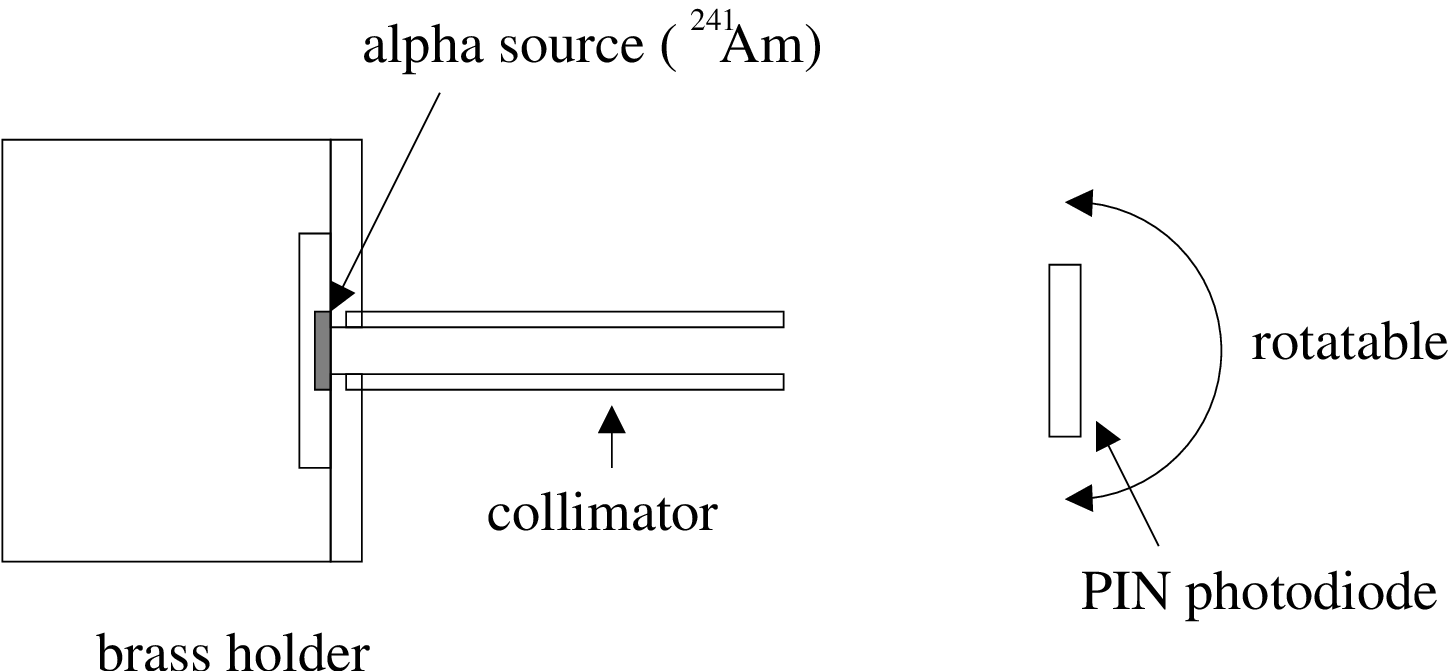}
\caption{Schematic view of the apparatus}
\label{setup1}
\end{center}
\end{figure}

\begin{figure}[p]
\begin{center}
\epsfxsize=8cm
\epsfbox{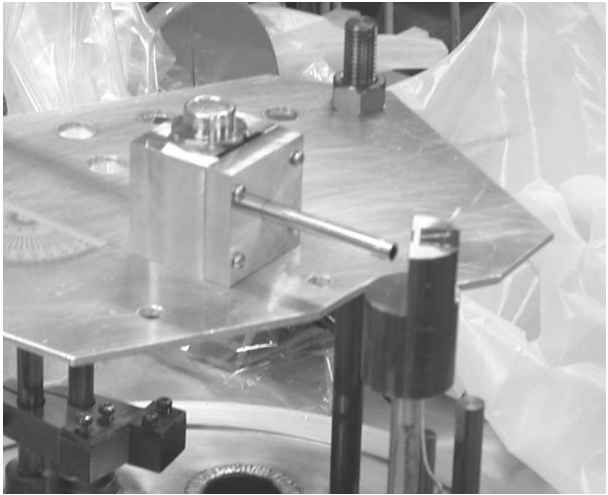}
\caption{Photograph of the experimental apparatus.
This device is put in vacuum.}
\label{pic}
\end{center}
\end{figure}

\begin{figure}[p]
\begin{center}
\epsfxsize=12cm
\epsfbox{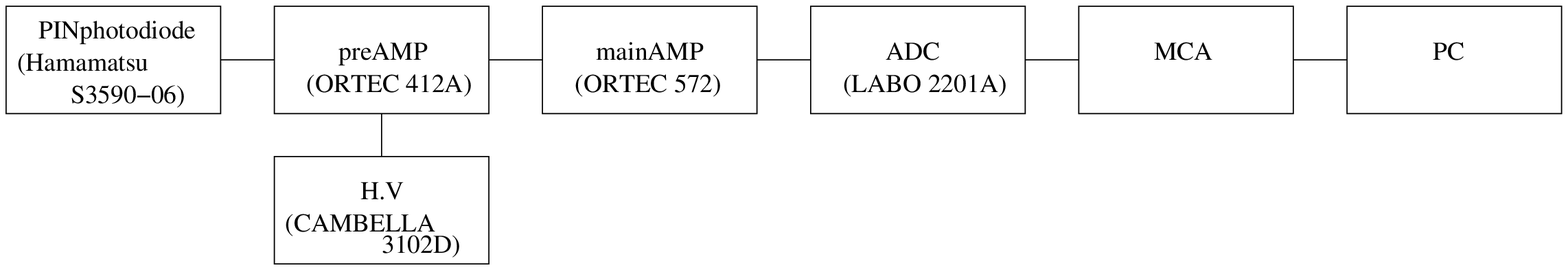}
\caption{Block diagram of the main system}
\label{setup2}
\end{center}
\end{figure}

The experimental setup is shown in Figs. $\ref{setup1}$, $\ref{pic}$
and $\ref{setup2}$. 
The experimental apparatus were put in a vacuum 
(pressure less than 10$^{-3}$ Pa).
We could rotate the PIN photodiode from the
outside of the vacuum to change the incidence angle.
The PIN photodiode was exposed to
5.486 MeV alpha-particles from the $^{241}$Am source installed in a brass 
holder.
The brass holder had a stainless steel pipe of $^{\phi}$4.3 mm inner diameter
and 50 mm length as a collimator.
The distance between the source and the detector is 60 mm. 

The PIN photodiode we used is Hamamatsu S3590-06.
It is 11$\times$11 mm$^2$
in size and 500 $\mu$m in thickness. This PIN photodiode is the same
type as ones used in the Tokyo Axion Helioscope \cite{tokyo1}\cite{tokyo2}.
It has high efficiency to x-rays of energies up to 10 keV,
energy range of x-rays to which solar axions convert in
magnetic fields.

\section{Result and Discussion}
\begin{figure}[p]
\begin{center}
\epsfxsize=6.8cm
\epsfbox{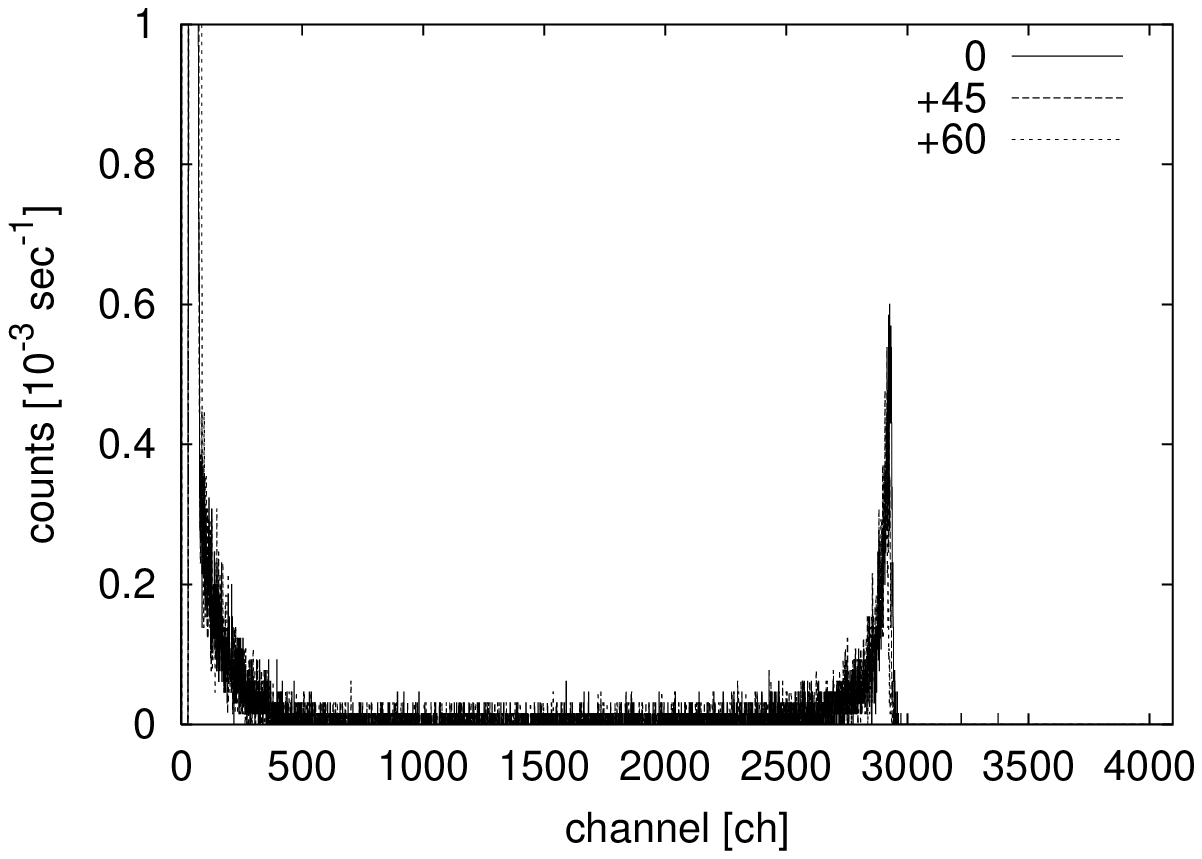}
\epsfxsize=6.8cm
\epsfbox{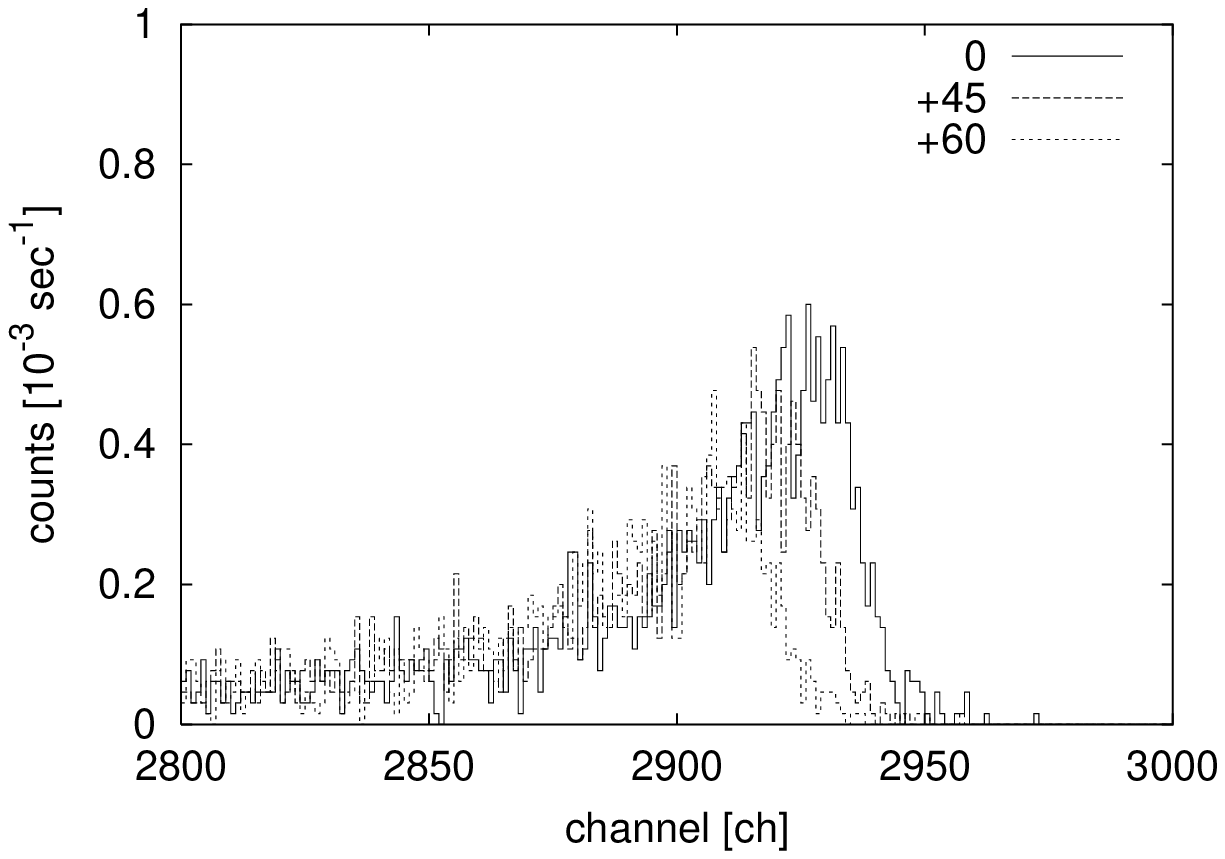}
\caption{Spectra of 3 typical incidence angles. Left figure shows spectra
in all channels and right is expanded spectra around the peak channels.}
\label{spectrum}
\end{center}
\end{figure}

We measured spectra for 8 incidence angles, $0^{\circ},\pm15^{\circ},
\pm30^{\circ},\pm45^{\circ}$, and $+60^{\circ}$ (Fig. \ref{spectrum}).
Each spectrum in Fig. \ref{spectrum} exhibits peaks of 5.486 MeV
around 2900 ch.
The alpha-particle peak channels shifted lower with increasing
incidence angles,
suggesting an existence of the insensitive surface layer.

We determined the peak channel $C(\theta)$ by fitting
a gaussian function to spectrum of the incidence angle $\theta$.
In the fitting, we use the upper half region of each peak, 
because the lower half contained the effect of the thickness of 
the $^{241}$Am source and
the spread of incidence angles of alpha-particles.

\begin{figure}[p]
\begin{center}
\epsfxsize=14cm
\epsfbox{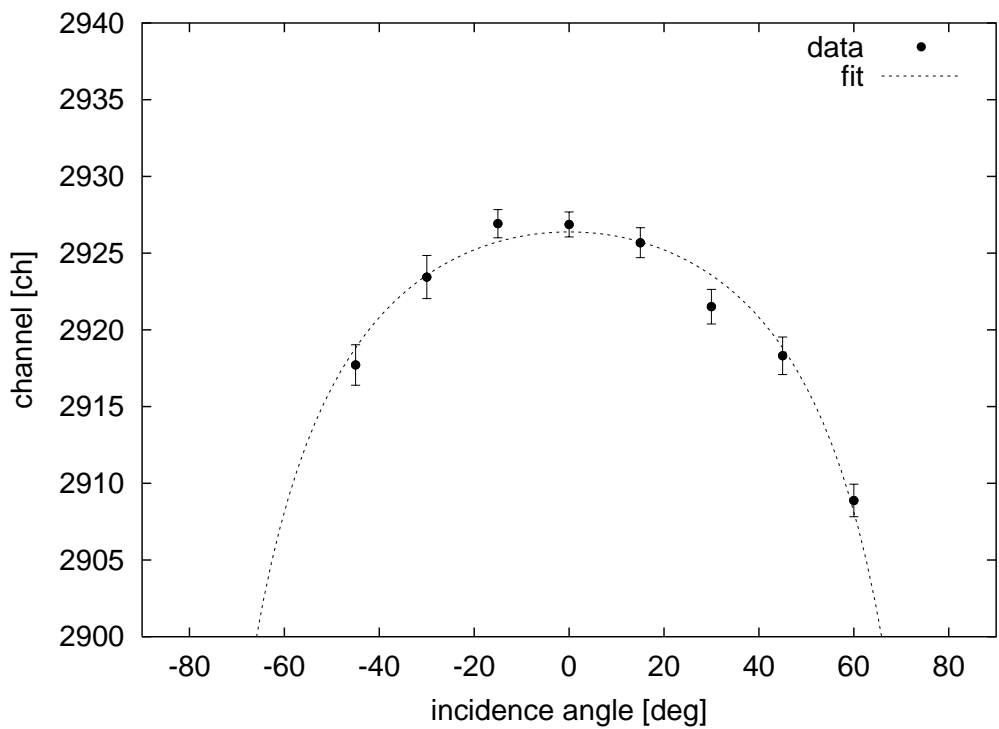}
\caption{Fitting of the alpha-particle spectrum of each incidence angle
with experimental data for Si. $\chi ^2/ {\rm ndf}=6.73/6$.}
\label{result}
\end{center}
\end{figure}

Fig. $\ref{result}$ shows the relation between the peak channels $C(\theta)$
and incidence angles $\theta$ together with the best fit to the data
by the equation,
\begin{equation}
C(\theta)=\frac{E_0}{\alpha} + \beta - \frac{{\rm d} E}{{\rm d} x} \cdot 
\frac{d}{\cos \theta} \cdot \frac{1}{\alpha},
\end{equation}
where $d$ is the thickness of the insensitive layer of the PIN photodiode,
$E_0$ is the energy of alpha-particles of $^{241}$Am,
$\frac{{\rm d}E}{{\rm d}x}$ is the stopping power, the insensitive
layer for the alpha-particles,
and $\alpha$ and $\beta$ are parameters of the relation between
channels and energies;
\begin{equation}
E \;{\rm [keV]}= \alpha (C\;{\rm [ch]} -\beta).
\end{equation}
Since we do not know the percentage of Si and SiO$_2$ in the insensitive
layer,
we calculated the thickness for the case it consisted of Si and the case
of SiO$_2$.
The stopping power of SiO$_2$ is a little larger than that of Si,
therefore, the estimate value of the thickness is a little smaller.
In the fitting, $d$ and $\beta$ were left free and $\alpha$ was fixed at
2.24.
This value was determined by calibration with
alpha-particles from $^{241}$Am in vacuum and 
decelerated alpha-particles from the same source in the air.
Thereby we got $d =0.31 \pm 0.02 \; {\rm \mu m}$ 
with $\chi ^2/ {\rm ndf}= 6.73 /6$ for Si
and 13\% smaller for SiO$_2$.
At worst, in the case that the insensitive layer consisted of Si,
the thickness is less than $0.35 \; {\rm \mu m}$ at 95\% confidence level.
This is nearly $\frac{1}{20}$ of the value we conservatively
assumed before \cite{tokyo1}.

\begin{figure}[p]
\begin{center}
\epsfxsize=14cm
\epsfbox{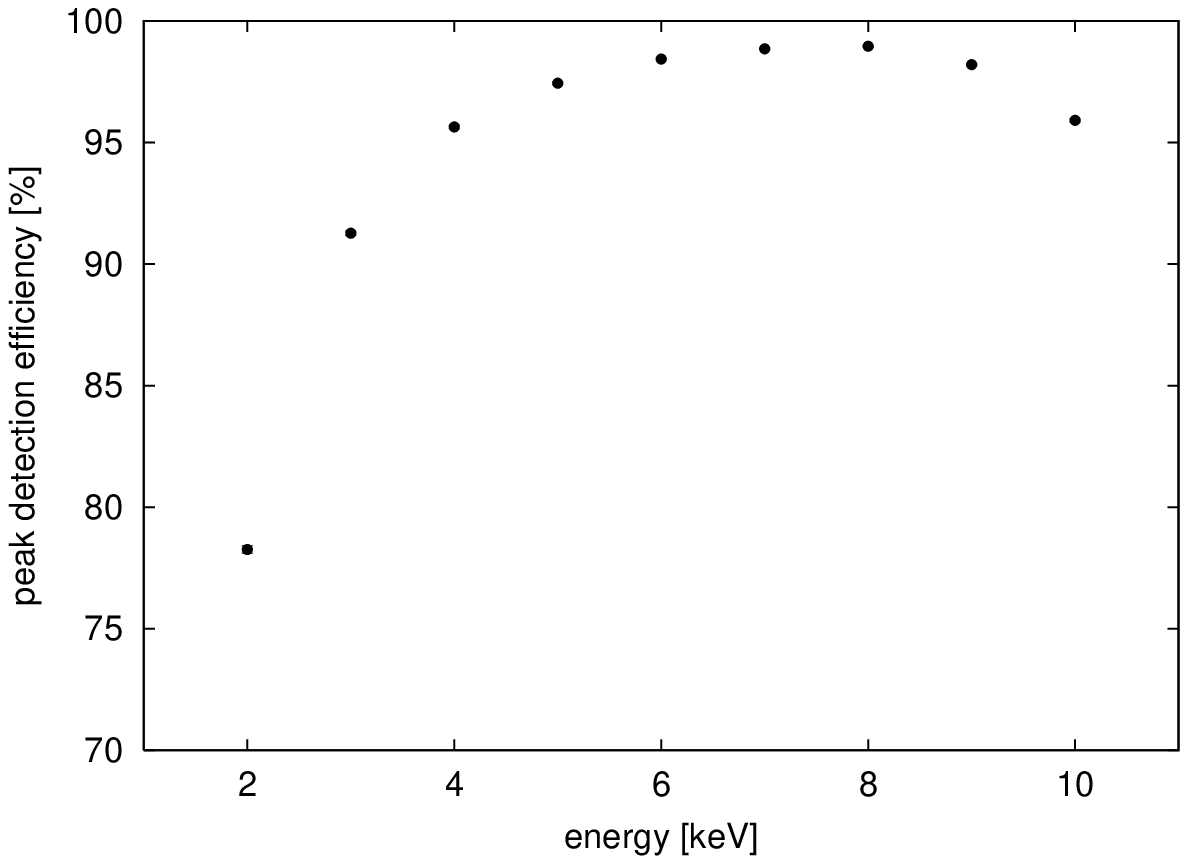}
\caption{Simulation result of peak detection efficiency of 
PIN photodiode for x-rays of various energies.}
\label{simulation}
\end{center}
\end{figure}

If the insensitive layer were relatively thick as we supposed before,
then the detection efficiency for the x-ray would be largely affected by
absorption in it.
However, it is not the case because the insensitive layer is thin
enough, less than 15\% of the absorption length for 5 keV x-rays.
However, one still has to take into account loss of energy caused by
escape of x-rays and Auger electrons from the detector following
photoelectric absorption of the incoming x-ray.
A Monte Carlo simulation was performed to estimate the peak detection
efficiency for x-rays of energy range between 2 keV and 10 keV
 with the GEANT4 program \cite{geant4}.
In the simulation
we assumed that the insensitive layer consists of Si.
The results are shown in Fig. \ref{simulation}.
The errors for the estimation are too small to show on the data points.
The estimated peak detection efficiency is more than 95\%
in the energy range between 4 and 10 keV.
The decreasing efficiency toward low energy is due to the absorption
in the insensitive layer and lower efficiency above 9 keV
is caused by the insufficiency of the thickness of the depletion layer.
Even if we simulated on the assumption that the insensitive layer 
consisted of SiO$_2$,
the result did not change significantly.

\section{Conclusion}
We showed the thickness of an insensitive layer of a PIN photodiode,
S3590-06 produced by Hamamatsu,
is $0.31 \pm 0.02 \; {\rm \mu m}$ for Si or
13\% smaller for SiO$_2$.
Therefore, the upper limit of the thickness is $0.35 \;{\rm \mu m}$
at 95\% confidence level in the case of the insensitive layer
consisted entirely of Si.
We calculated the peak efficiency of a PIN photodiode for x-rays 
in consideration of
the thickness of the insensitive layer
and escape of x-rays and Auger electrons from the detector.
This result showed the efficiency as more than 95\% for 4-10 keV x-rays,
which correspond to the energy range observed in the Tokyo Axion
Helioscope experiment \cite{tokyo1} \cite{tokyo2}.


\begin{thebibliography}{00}
\bibitem{tokyo1}
S. Moriyama, {\it et al.}, Phys. Lett. B {\bf 434} (1998) 147.
\bibitem{tokyo2}
Y. Inoue, {\it et al.}, Phys. Lett. B {\bf 536} (2002) 18.
\bibitem{pin}
Y. Inoue, {\it et al.}, Nucl. Instr. Methods in Phys. Research A {\bf 368}
(1996) 556.
\bibitem{geant4}
http://geant4.web.cern.ch/geant4/
\end{thebibliography}
\end{document}